# Augmenting endometriosis analysis from ultrasound data with deep learning

Adrian Balica*[,a] and Jennifer Dai*[,b] and Kayla Piiwaa[b] and Xiao Qi[c] and Ashlee N. Green[a] and Nancy Phillips[a] and Susan Egan[a] and Ilker Hacihaliloglu[d]

[a] Department of Obstetrics and Gynecology, Robert Wood Johnson University Hospital, USA
[b] Robert Wood Johnson Medical School, Rutgers University, USA
[c] Department of Electrical and Computer Engineering, Rutgers University, USA
[d] Department of Radiology, Department of Medicine, University of British Columbia, Canada

**ABSTRACT**

Endometriosis is a non-malignant disorder that affects 176 million women globally. Diagnostic delays result in severe dysmenorrhea, dyspareunia, chronic pelvic pain, and infertility. Therefore, there is a significant need to diagnose patients at an early stage. Our objective in this work is to investigate the potential of deep learning methods to classify endometriosis from ultrasound data. Retrospective data from 100 subjects were collected at the Rutgers Robert Wood Johnson University Hospital (New Brunswick, NJ, USA). Endometriosis was diagnosed via laparoscopy or laparotomy. We designed and trained five different deep learning methods (Xception, Inception-V4, ResNet50, DenseNet, and EfficientNetB2) for the classification of endometriosis from ultrasound data. Using 5-fold cross-validation study we achieved an average area under the receiver operator curve (AUC) of 0.85 and 0.90 respectively for the two evaluation studies.

## 1. INTRODUCTION

Endometriosis is a non-malignant disorder that affects 10-15% of American women and 176 million women globally[1,2]. Symptoms of endometriosis include severe dysmenorrhea, dyspareunia, and chronic pelvic pain[2]. Without treatment, endometriosis can lead to infertility; thus, there is a significant need to diagnose patients early[3,4]. Despite this, the mean time to diagnosis is seven years[5]. While the current gold standard for diagnosis is laparoscopy with tissue diagnosis, there is not yet a reliable, nonsurgical, diagnostic method[5]. Ultrasound could aid in clinical diagnosis and provide a safer, non-invasive, and more cost-effective diagnostic option[6]. However, ultrasound data is characterized by high levels of noise, imaging artifacts, and variability during data collection (manual data collection). Furthermore, identification of early lesions in qualitative ultrasound is challenging[5,6].

Deep learning, a subfield of machine learning, has recently been successfully deployed for processing various medical data to support clinical decision-making for a range of diseases[7,8], including endometriosis[9-12]. In Hu et. al. (2019)[9], a deep learning method (based on previously proposed two well-known neural network architectures VGG[13] U-net[14]) was proposed for segmentation and thickness measurement of the endometrium. In Guerroro et. al. (2021)[10], k-nearest neighbors algorithm (k-NN), Naive Bayes, Neural Networks (NNET-neuralnet), Support Vector Machine (SVM), Decision Tree, Random Forest, and Logistic Regression models were investigated. The highest reported area under the receiver operating curve (AUC) was 81.73% for NNET-neuralnet using single-fold validation. In Yang M et. al. (2021)[11], a traditional deep learning method[13] was extended for classifying deep pelvic endometriosis (DPE). The reported single-fold classification accuracy was 96.5%, however, quantitative results for AUC were not reported. The authors also did not mention the size of the training and validations set nor provided information if the same patient scans were used for training and testing. In Maicas et. al. (2021)[12], ResNet with spatiotemporal information[15], for the classification of the sliding sign for the diagnosis of the pouch of Douglas (POD) obliteration, was investigated. Single-fold validation achieved an average accuracy of 88.8% and AUC 96.5%. The evaluation dataset included 86.2% positive and 13.8% negative POD group. Data augmentation to balance the class distribution was not performed.

*These authors contributed equally to this work. kap421@rwjms.rutgers.edu, ilker.hacihaliloglu@ubc.ca

The primary objective of this pilot study was to evaluate the feasibility of using deep learning methods to diagnose the presence of endometriosis from ultrasound data. Specifically, we optimize the design of the five most widely used deep learning methods for processing medical image data, Xception[26], Inception-V4[27], ResNet50[16], DenseNet[18] and Efficient-NetB2[17], for improved classification endometriosis from clinical ultrasound data.

## 2. METHODS AND MATERIALS

### 2.1 Data collection

The data was collected as part of an IRB-approved retrospective chart review, so consent was waived. In total 100 subjects were included in the study: 50 subjects with no endometrial findings on ultrasound, and 50 with diagnosed endometriosis who were seen at Rutgers University Robert Wood Johnson University Hospital (New Brunswick, NJ, USA) between 2017-2020. Subjects in the endometriosis group were identified with a positive diagnosis of endometriosis from the medical record. All the subjects 21 years of age or older and diagnosed with endometriosis via laparoscopy or laparotomy and assessed via transabdominal and transvaginal ultrasounds were included in the data collection. Out of the 50 endometriosis subjects, 34 subjects had laparoscopy or laparotomy procedures at Rutgers, while 16 had a documented procedure in the electronic medical records (EMR) at another site. A total of 174 subjects > 21 years old from 2018-2020 with normal ultrasound findings evaluated at Rutgers were identified for inclusion in the normal group. The normal group (no endometriosis group) was composed of patients with at least one symptom suggestive of endometriosis but without previous treatment for endometriosis and ultrasound findings not suggestive of endometriosis. From the normal group, 2 subjects were removed because they were later diagnosed with endometriosis. Subjects who had undergone hysterectomies, oophorectomies, or had ultrasounds outside of IRB-approved dates were also excluded from both groups.

Ultrasound protocol at Rutgers included the following. Transabdominal ultrasound was performed with a Voluson E8 scanner equipped with an abdominal 4C-D 2–5-MHz transducer (GE Healthcare, Milwaukee, WI). The ultrasound data were collected by an AIUM-certified ultrasound technician. For each subject, all ultrasound images within the IRB-approved time frame were downloaded. These ultrasound images were stratified by subject for analysis. In total, we had 1,788 ultrasound scans from the endometriosis group and 812 scans from the normal group. All the subjects also had video ultrasound recordings available as part of their scanning protocol, and these were included in the evaluation study as well.

### 2.2 Data Pre-processing

Ultrasound images were processed to standardize input into the deep learning algorithm. The ultrasound image intensities were rescaled to a certain size (0-255). To normalize the image size, we cropped a centered square region of interest, with the size of the smallest image axis. Images were resized to 512×512 pixels to be processed by the CNN architecture. image zooming by a factor uniformly sampled from [0.95, 1.05], translation in 4 directions by a factor uniformly sampled from +- 5% of height and width, and rotation by a factor uniformly sampled from [-15, 15] degrees. After data augmentation, the training dataset consisted of 3,412 images (1,788 images from endometriosis patients, and 1,624 images from normal patients). We referred to the test data used in this analysis as 'Test 1' data. Five-fold cross-validation was used during evaluation. The data partition for each fold was divided into 60% of images used for training, 20% for validation, and 20% for testing. Test data did not include augmented data to ensure accuracy metrics were calculated on previously unseen and unprocessed images. During data partitioning, all images from the same patient appeared in the same partition. At the end of this process, 'Test 1' data had 520 images (357 images from the endometriosis group and 163 images from the normal group). An additional 41,186 images including 26,250 positive images (diagnosed with endometriosis) and 14,936 negative images (no endometriosis diagnosis) were extracted from video data and were referred to as 'Test 2' data. We also ensured that the 'Test 2' data did not have scans used for training the deep learning methods.

### 2.3 Deep Learning Methods

As part of this work, we have evaluated three different convolutional neural network (CNN) architectures: Xception[26], Inception-V4[27], ResNet50[16] DenseNet[18] and EfficientNetB2[17]. These are the most widely investigated CNN models for classifying medical image data[7, 8]. Each of these architectures was extended by using one fully connected layer with half the length of the input features with rectified linear unit (ReLU) activation[23] and a final classification layer of length one with sigmoid activation. Dropout layers with a dropout coefficient of 0.2 were interleaved between the fully connected layers. The model weights were pre-trained using the ImageNet[10] database. The loss function employed was binary cross-entropy loss[7,8]. We balanced the binary cross-entropy loss of positive and negative samples within each batch following[19]. The experiments were implemented in Python using the Pytorch framework[22]. The networks were trained using an AdamW optimizer[21] with a learning rate schedule that decreased the learning rate between $10^{-4}$ for the first epochs to $10^{-6}$ for the last ones. The mini-batch size was set to 16. The learning rate was reduced by a factor of 10 if the validation area under the receiver operating curve (AUC) did not improve for 10 epochs. The training was performed for 60 epochs.

## 3. RESULTS

The Densenet CNN algorithm successfully detected the presence of endometriosis on ultrasound with an AUC of 90% and an accuracy of 80% using heterogeneous patterns of endometriosis. This algorithm has the potential to be extremely useful in the clinic, as it can take as input a patient's ultrasound, and suggest with 90% probability and 80% accuracy whether the patient has endometriosis. This tool, combined with the patient's clinical presentation, has the potential to advance our current capability to diagnose endometriosis.

Other algorithms were also evaluated. Table 1 shows the average accuracy of the 5-fold cross-validation on the 'Test 1' and 'Test 2' datasets. All investigated network designs achieved an AUC above 80% and an accuracy above 70%. For 'Test 1' data, the original dataset with data augmentation, Densenet outperformed the other networks (p<0.05 for paired t-test, Figure 1, Table 1). For 'Test 2' data, the AUC and accuracy results improved to 90% and 84% respectively (Figure 2, Table 1). The highest AUC and accuracy were again obtained using the Densenet network architecture.

**Table 1.** Precision, Recall, F1-Score, Accuracy, and AUC values for the designed and optimized deep learning methods. Average AUC and ACC values across five folds. Results show the average values across five folds.

| Model | Precision | | Recall | | F1-Scores | | Top-1(%) Accuracy | | AUC | |
|---|---|---|---|---|---|---|---|---|---|---|
| | Test-1 | Test-2 | Test-1 | Test-2 | Test-1 | Test-2 | Test-1 | Test-2 | Test-1 | Test-2 |
| XCeption | 0.74 | 0.78 | 0.75 | 0.78 | 0.74 | 0.78 | 0.75 | 0.78 | 0.816 | 0.861 |
| InceptionResnetv2 (Inceptionv4) | 0.76 | 0.79 | 0.76 | 0.79 | 0.76 | 0.79 | 0.76 | 0.79 | 0.820 | 0.868 |
| ResNet50 | 0.74 | 0.80 | 0.74 | 0.78 | 0.74 | 0.76 | 0.74 | 0.78 | 0.822 | 0.884 |
| EfficientnetB2 | 0.74 | 0.81 | 0.75 | 0.79 | 0.74 | 0.78 | 0.75 | 0.79 | 0.828 | 0.876 |
| Densenet121 | 0.77 | 0.81 | 0.76 | 0.80 | 0.76 | 0.79 | 0.76 | 0.80 | 0.846 | 0.900 |

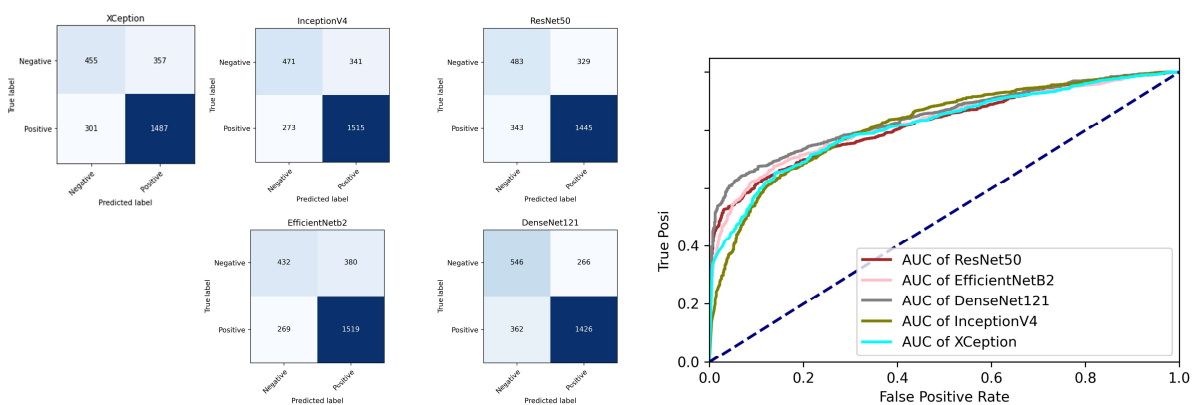

**Fig. 1.** AUC plots and confusion matrix results for all the investigated network architectures. Quantitative results were obtained using 'Test 1' data, which had 2,600 ultrasound images in total after data augmentation for 5-fold cross-validation.

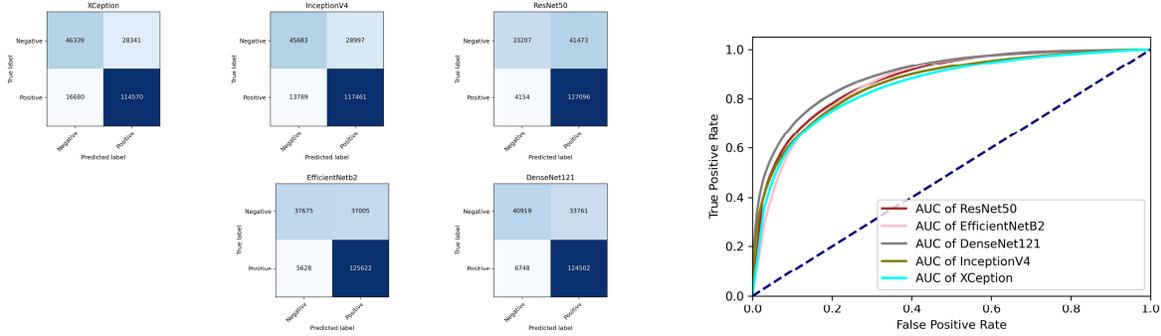

**Fig. 2.** AUC plots and confusion matrix results for all the investigated network architectures. Quantitative results were obtained using 'Test 2' data which had 205,930 ultrasound images in total for 5-fold cross validation. Data augmentation was not performed for this data set.

## 4. DISCUSSIONS AND CONCLUSIONS

Early detection of endometriosis is crucial to prevent adverse outcomes such as infertility. Ultrasound, due to being non-invasive, safe, and cost-effective (compared to MRI), is a promising imaging modality providing a safer alternative to gold standard invasive laparoscopy-based diagnosis. However, high levels of noise, imaging, artifacts, and being a user-operated imaging modality makes the diagnosis of endometriosis from ultrasound data challenging. In this work, we have investigated the potential of artificial intelligence, based on deep learning, methods to diagnose endometriosis from ultrasound data. Five different deep-learning methods were designed and optimized to classify endometriosis from ultrasound data. We have achieved close to 0.9 average AUC and 80% accuracy in a 5-fold cross-validation study meaning that given an ultrasound, we can predict with 90% probability and 80% accuracy that the patient has endometriosis. To balance the dataset distribution we have performed data augmentation. Our test data did not include any augmented data to ensure that the quantitative evaluation was not influenced by the augmented data. Our study is an initial successful pilot study to determine the presence of endometriosis using deep learning. Strengths include being the first study to evaluate the classification of heterogeneous patterns of endometriosis on patients with a surgical diagnosis of endometriosis using CNN architectures, making our study most applicable to the clinic. We also designed and optimized the most widely used CNN architectures for medical image classification as a proof-of-concept study. The present study is only a preliminary attempt to integrate deep learning into the clinical decision-making process for improved management of endometriosis. Several challenges remain to be addressed which are discussed in the next section.

Our initial pilot study included scans collected only from 100 patients retrospectively. This is not a significant amount of data to harness the real power of deep learning methods. Clinical data collection is currently ongoing in our institute. During the data collection, patients with fibroids and other reproductive abnormalities were included, which may have affected the regions the algorithm uses to classify endometriosis. However, women with endometriosis commonly present with these comorbidities, making our analysis applicable in the clinic. In Hacihaliloglu and Balica (2020)[24], we have shown that ultrasound tissue representation of endometriosis can be improved by using local phase-based image enhancement methods. Local phase tissue signatures are intensity invariant and are not affected by the ultrasound transducer operating frequency, ultrasound machine settings, and the body mass index of the patient. Figure 3 shows the enhancement results obtained from our local phase-based image processing method. In Che et al (2021)[25], we have shown that by using the local phase image features the success of deep learning methods for diagnosing liver disease from ultrasound data can be improved compared to the most widely used deep learning methods such as the ones investigated in this work. We did not include such an analysis in this work. Our future work will involve the incorporation of the local phase images for the design of multi-feature deep learning methods for improving classification performance.

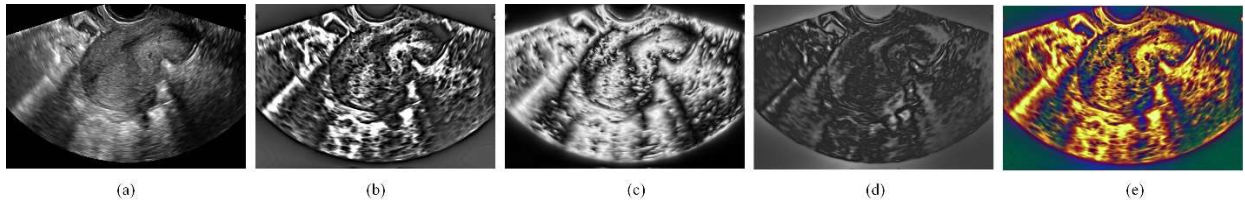

**Fig. 3.** Local phase-based image enhancement results. (a) B-mode ultrasound image with laparoscopy confirmed endometriosis. (b-d) three different local phase images. (e) Multi-feature image obtained using the combination of the local phase images shown in (b-d).

Many clinical applications have successfully analyzed ultrasound images using convolutional neural networks (CNN). With this pilot study, we have shown that traditional CNN architectures can classify the presence of endometriosis from ultrasound data. Future work will include increasing our patient population and evaluating multi-feature CNN architectures[25] for improved diagnostic accuracy.

## 5. ACKNOWLEDGEMENTS

The authors thank Allison Cabinian (Rutgers Women Health Institute, New Brunswick, New Jersey, USA) for her contribution.